\documentclass[epj]{webofc}
\usepackage[varg]{txfonts}   
\wocname{EPJ Web of Conferences}
\woctitle{CONF12}
\usepackage{color}

\usepackage[normalem]{ulem}  

\newcommand{\non}{\nonumber\\}

\newcommand{\be}{\begin{equation}}
\newcommand{\ee}{\end{equation}}
\newcommand{\bea}{\begin{eqnarray}}
\newcommand{\eea}{\end{eqnarray}}

\begin{document}
\selectlanguage{english}
\title{Mixing of charged and neutral Bose condensates at nonzero\\ temperature and magnetic field}
%
%

\author{Alexander Haber\inst{1,2}\fnsep\thanks{\email{ahaber@hep.itp.tuwien.ac.at}} \and
        Andreas Schmitt\inst{2}\fnsep\thanks{\email{a.schmitt@soton.ac.uk}}
}

\institute{Institut f\"{u}r Theoretische Physik, Technische Universit\"{a}t Wien, 1040 Vienna, Austria
\and
          Mathematical Sciences and STAG Research Centre, University of Southampton, Southampton SO17 1BJ, United Kingdom
}

\abstract{It is expected that in the interior of compact stars a proton superconductor coexists with and couples to a neutron superfluid. Starting from a field-theoretical model 
for two complex scalar fields -- one of which is electrically charged -- we derive a Ginzburg-Landau potential which includes entrainment between the two fluids and temperature effects from thermal excitations of the 
two scalar fields and the gauge field. 
The Ginzburg-Landau description is then used for an analysis of the phase structure in the presence of an external magnetic field. 
In particular, we study the effect of the superfluid on the flux tube phase by computing the various critical magnetic fields and deriving an approximation for the flux tube interaction. As a result, we point out differences to the naive expectations from an isolated superconductor, for instance the existence of a first-order flux tube 
onset, resulting in a more complicated phase structure in the region between type-I and type-II superconductivity.
   }
\maketitle
\section{Introduction and main results}
\label{intro}

\begin{figure}[t]
\centering
\includegraphics[width=\textwidth]{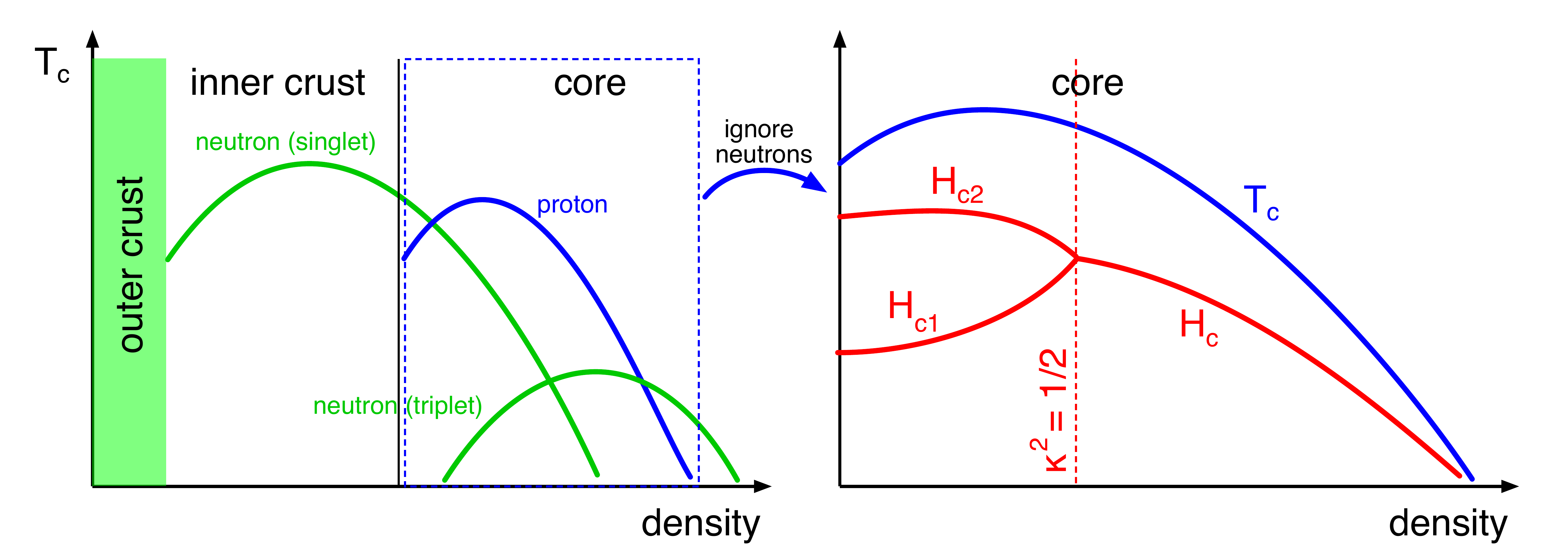}
\caption{Schematic view of the interior of a compact star. (Reproduced with modifications from Refs.\ \cite{Glampedakis:2010sk,Graber:2016imq}.) Left panel: critical temperatures for proton superconductivity 
and neutron superfluidity, where Cooper pairing can occur in the ${}^1S_0$ (singlet) or ${}^3P_2$ (triplet) channels. Outer and inner crust contain a lattice of ions. In the inner crust, a neutron superfluid is immersed in this 
lattice. In the core, where the density exceeds nuclear saturation density, neutron superfluidity is expected to coexist with a proton superconductor. Right panel: critical magnetic fields for the proton superconductor, 
as expected naively from a system without a coexisting superfluid (the present work points out deviations from this expectation, see Figs.\ \ref{fig:TcHc} and \ref{fig:HHH}). 
An array of magnetic flux tubes exists between the 
critical magnetic fields $H_{c1}$ and $H_{c2}$ ("type-II superconductor"), and below a certain value of the Ginzburg-Landau parameter $\kappa$, no flux tubes are expected for any external magnetic field 
("type-I superconductor"). The details of all curves shown here are poorly known at large densities: 
for instance, it is not clear whether singlet and triplet neutron pairing indeed coexist in a certain density regime, whether the density becomes large enough to indeed realize a type-I superconductor etc.\ (let alone
the possiblility that the transition to a quark matter phase cuts off the shown nuclear matter phases somewhere in the core). 
} 
\label{fig:schematic} 
\end{figure}

In the dense environment of compact stars, nucleons can form Cooper pairs, just like electrons in an ordinary superconductor \cite{bogol,1959NucPh..13..655M}. 
As a consequence, nuclear matter becomes superconducting and/or superfluid, depending 
on whether protons pair, or neutrons, or both (cross-pairing of neutrons with protons is strongly suppressed because of the large difference in Fermi momenta). In addition, compact stars can have large magnetic fields and can spin fast, resulting in a possible coexistence of magnetic flux tubes and superfluid vortices, not necessarily aligned with each other. The pairing gaps for both neutron and 
proton Cooper pairing depend on density (in a non-monotonic way) and become very small in certain density regions of the star \cite{Page:2013hxa}. Therefore, temperature effects become important because the critical temperature is of the same order of magnitude as the pairing gap and thus can become very small too. The same is true for the critical magnetic fields: if the pairing gap is small, only a small magnetic field is needed to either penetrate into the superconductor through flux tubes or to 
break superconductivity completely. A schematic view of this complicated system is shown in Fig.\ \ref{fig:schematic}. In this figure, as well as in the whole study, we ignore the rotation of the system, i.e., we do allow 
for magnetic flux tubes, but not for rotational vortices. 

\begin{figure}[t]
\centering
\hbox{\includegraphics[width=0.5\textwidth]{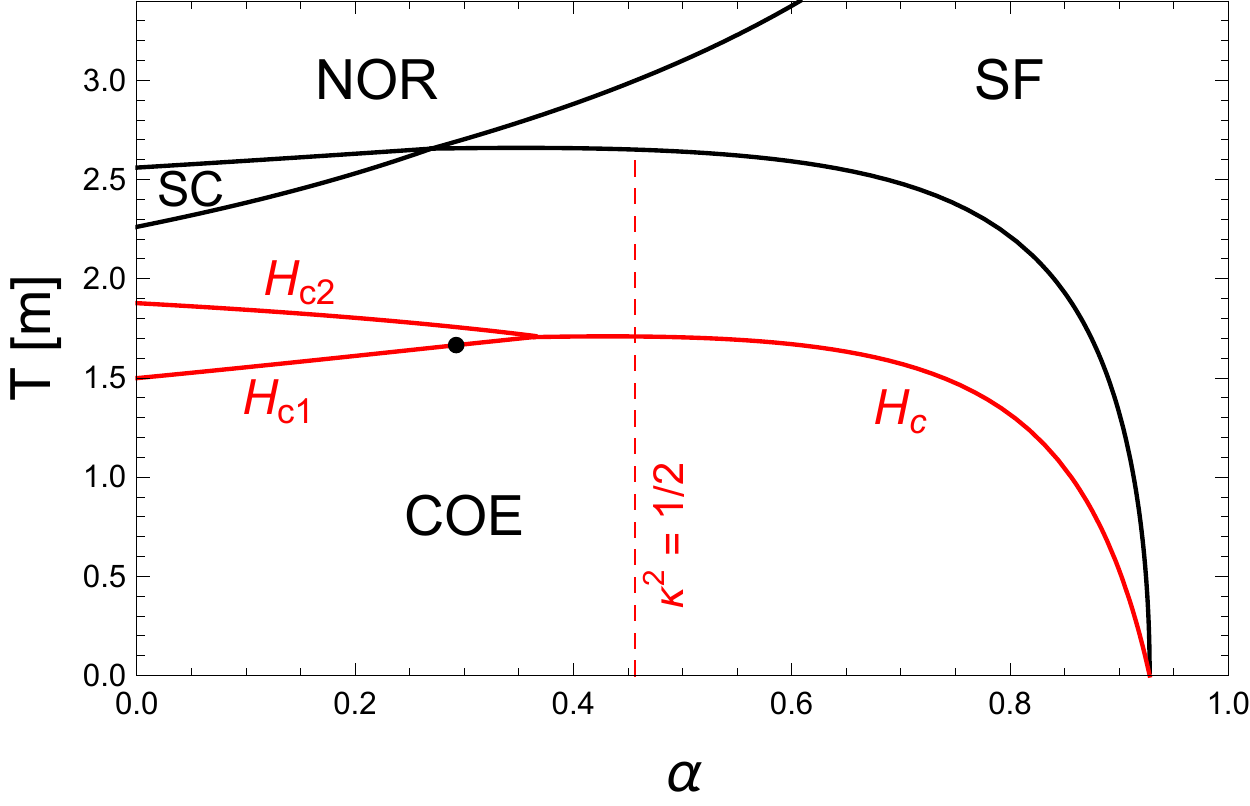}\includegraphics[width=0.5\textwidth]{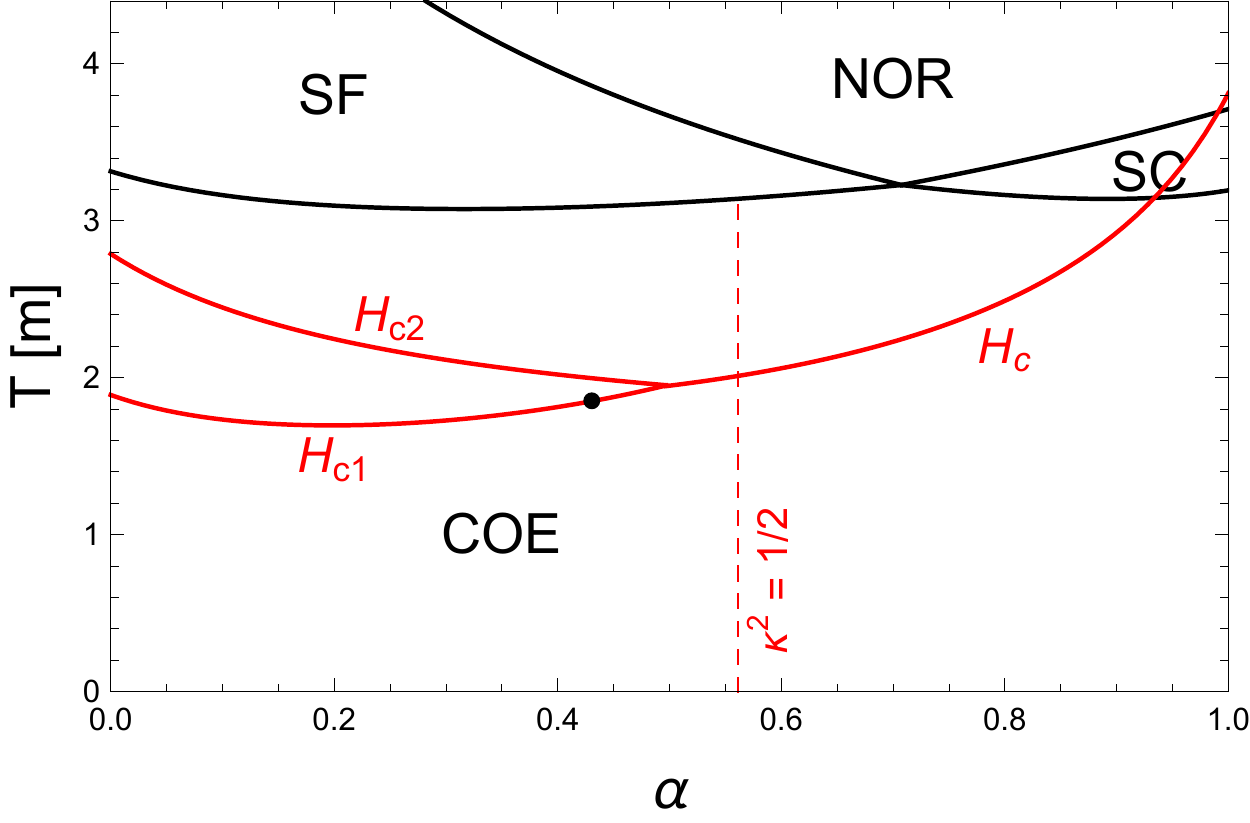}}
\caption{Critical temperatures and zero-temperature critical magnetic fields for a negative ($h<0$, left panel) and a positive ($h>0$, right panel) value of the non-entrainment coupling and a certain 
path through the two-dimensional parameter space spanned by the self-couplings $\lambda_1$, $\lambda_2$, parametrized by $\alpha$, for details see Fig.\ \ref{fig:l1l2}. Here we have set the entrainment coupling to zero. The (black) dot on the $H_{c1}$ curve marks the point where the second-order transition to the flux tube phase becomes first order, 
see Fig.\ \ref{fig:HHH} for a schematic view of this region and Sec.\ \ref{sec:inter} for more details. The critical magnetic fields are given in units of $m^2$ and are multiplied by $0.3$ (left) and $0.2$ (right) to fit into the plot. We have used the following abbreviations: 
COE = coexistence of superfluid and superconductor, SC = superconductor with vanishing neutral condensate, SF = superfluid with vanishing charged condensate, NOR = both condensates vanishing.} 
\label{fig:TcHc} 
\end{figure}
\begin{figure} [h]
\centering
\includegraphics[width=\textwidth]{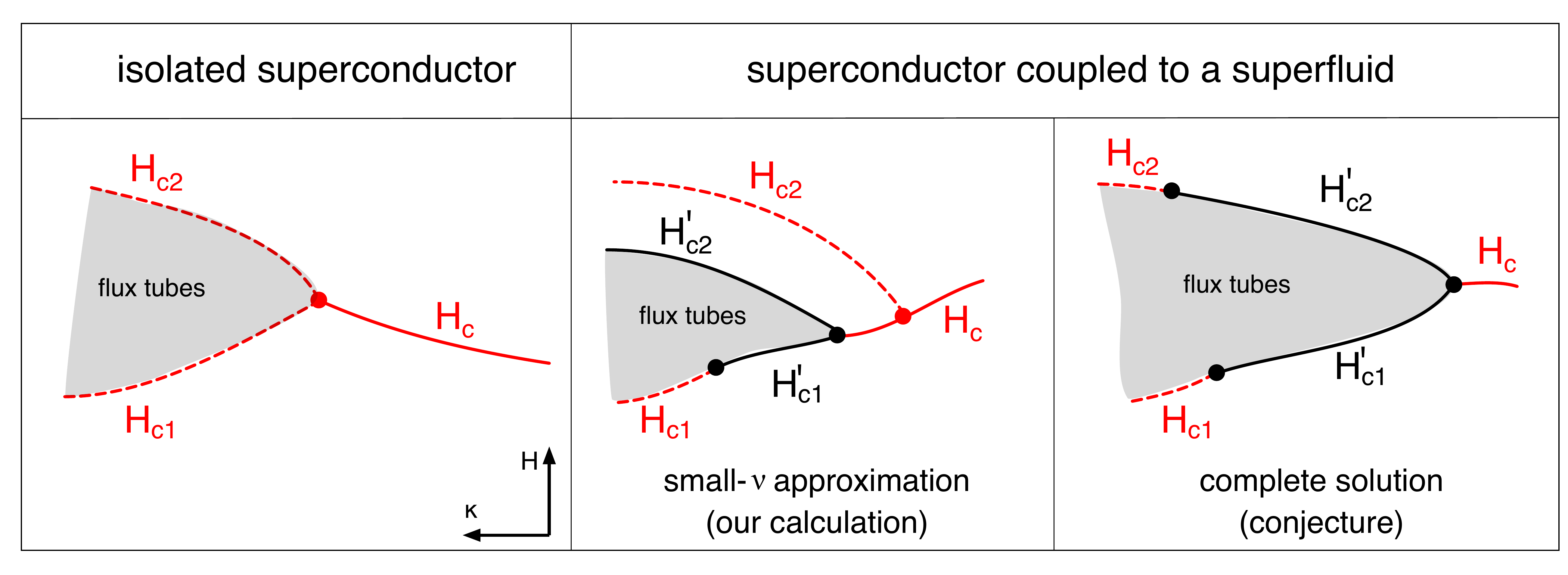}
\caption{Illustration of the type-I/type-II transition region for an isolated superonductor and a superconductor coupled to a superfluid. We show 
rigorously that the second-order transition at $H_{c1}$ turns into a first-order phase transition at $H_{c1}'$. Stretching our approximation of small flux tube area densities $\nu$
beyond its regime of validity, we also compute $H_{c2}'$, which yields the unphysical result $H_{c2}'<H_{c2}$. The simplest phase structure suggested by these results is
shown on the right-hand side, where the flux tube phase is bounded by first-order phase transitions close to the type-I/type-II transition region.}
\label{fig:HHH}
\end{figure}

In the present work, we consider a simplified version of this complicated system. 
We start from a relativistic model for two bosonic fields, one of them charged, include entrainment between them, and reduce it to a time-independent Ginzburg-Landau 
potential, similar to previous non-relativistic approaches \cite{AlparLangerSauls:1984,Alford:2007np}. Since we intend to apply the model to 
nonzero temperatures, we compute the thermal excitations and derive the temperature dependence of the Ginzburg-Landau coefficients, which,
in the presence of entrainment, has not been done previously.  
We present the main results in Figs.\ \ref{fig:TcHc} and \ref{fig:HHH}, with derivations and details in the subsequent sections. 
Instead of fitting our parameters to poorly known properties of dense nuclear matter, we have tried to keep our study as general as possible. In doing so we 
have to deal with an unwieldy multi-dimensional parameter space, necessarily making some assumptions and simplifications.  
As a result of our general approach, only the topology, not the details, in the phase diagrams in Figs.\ \ref{fig:TcHc} and \ref{fig:HHH} are relevant. Nevertheless, we obtain some interesting insight when we compare the results of our calculation with the schematically 
shown expectation of Fig.\ \ref{fig:schematic}. First of all, our approach predicts, for a given parameter set, the phase structure for all temperatures and magnetic fields. For example, if we start from a phase 
where superfluid and superconductor coexist, increasing the temperature may lead us to a pure superfluid phase or a pure superconducting phase, depending on the values of the self- and cross-couplings. This 
fully consistent treatment of all possible phases is usually simplified in the calculations that underly the sketch in Fig.\ \ref{fig:schematic}. Regarding the critical magnetic fields, the most obvious conclusion is that the change from type-I to type-II superconductivity does not occur at the usual value for the Ginzburg-Landau parameter $\kappa^2=1/2$, as 
already noticed for instance in Ref.\ \cite{Buckley:2003zf}. However, we find a further discrepancy to the case of an isolated superconductor when we have a closer look at the 
transition region: the three critical magnetic fields $H_c$, $H_{c1}$, and $H_{c2}$ do not intersect in one point, but in three distinct points. And the point 
where the interaction between flux tubes changes from being attractive to repulsive at large distances -- usually an indication for the type-I/type-II transition -- 
coincides with neither of the three intersections. 
This transition region will be discussed in more detail in Sec.\ \ref{sec:inter}, including a general expression for the flux tube interaction at large distances. However, eventually, a full numerical evaluation of the flux tube phase is necessary to complete this region of the phase diagram. We leave such a study 
for the future. But, based on our results, the shape of the complete result can be conjectured, see Fig.\ \ref{fig:HHH}.


Our model 
allows for further generalizations and applications to different systems. First of all, we could consider topological defects of the neutral condensate instead of the charged condensate. This scenario, 
superfluid vortices getting magnetized through entrainment of a charged fluid, is also relevant for dense nuclear matter inside a compact star \cite{Alpar:1984zz,1995ApJ...447..305S}, and 
it would be interesting to see whether our numerical calculation of the vortex profile, taking into account the coupling between the fluids from a 
"microscopic" point of view, agrees with or possibly goes beyond the traditional approaches in the literature. It is also possible to apply the model in a modified form 
to the color-flavor locked quark matter phase with kaon condensation \cite{Alford:2007xm} and see whether and how color magnetic flux tubes \cite{Alford:2010qf} are affected by the presence of the kaon condensate. In experiments with ultra-cold atoms, mixtures of charged and neutral gases seem impossible to create because atoms are neutral. However, on the one hand, Bose-Fermi mixtures with two neutral condensates have been created \cite{2014Sci...345.1035F}, and, on the other hand, flux tubes in a "synthetic magnetic field" have been observed in a single-component system \cite{2009Natur.462..628L}. It thus seems conceivable to create 
two-component systems where only one component couples to the synthetic magnetic field and for which our model might be suitable. Furthermore, the model obviously can be used in a completely ungauged version, describing two superfluids. This has been done to investigate hydrodynamical instabilities
\cite{Haber:2015exa}, which is of phenomenological interest as a mechanism for pulsar glitches, but also in ultracold atomic Bose-Fermi systems \cite{2015PhRvL.115z5303D}. Or, both fields can be charged: two-component superconductors 
can be realized for instance in two-band superconductors \cite{Carlstrom:2010wn}, in liquid metallic hydrogen \cite{babaev2004superconductor}, or possibly in compact stars if
charged hyperons form Cooper pairs \cite{Gusakov:2009kc}.

\section{Ginzburg-Landau potential and critical magnetic fields }

We  consider the Ginzburg-Landau free energy
\be
F = \int d^3\vec{x} \, U(\vec{x}) \, , 
\ee
with the potential 
\bea \label{Ux}
U(\vec{x}) &=& \frac{(\nabla\rho_1)^2}{2}+\frac{(\nabla\rho_2)^2}{2}- \frac{\mu_1^2-(\nabla\psi_1-q\vec{A})^2-m_{1,T}^2}{2}\rho_1^2-
\frac{\mu_2^2-m_{2,T}^2}{2}\rho_2^2+\frac{\lambda_1}{4}\rho_1^4 +\frac{\lambda_2}{4}\rho_2^4 \non[2ex]
&&-\frac{h_T}{2}\rho_1^2\rho_2^2 -\frac{G}{2}\rho_1\rho_2\nabla\rho_1\cdot\nabla\rho_2+\frac{B^2}{8\pi} \, ,
\eea
where
\be\label{mT}
m_{1,T}^2 = m_1^2 + \frac{2\lambda_1-h+6\pi q^2}{6}T^2 \, , \quad m_{2,T}^2 = m_2^2 + \frac{2\lambda_2-h}{6}T^2\, , \qquad h_T = h\left(1+\frac{GT^2}{6}\right) \, .
\ee
This potential, in particular its temperature dependence, can be derived from an underlying Lagrangian for two complex scalar fields $\varphi_1$, $\varphi_2$, 
whose condensates are parameterized by $\langle\varphi_i\rangle = \rho_i e^{i\psi_i}/\sqrt{2}$, for details of the derivation see Ref.\ \cite{Haber:2017kth}.
One of the fields, here field 1, carries charge $q$ and couples to the gauge field $\vec{A}$, with the magnetic field 
$\vec{B}=\nabla\times \vec{A}$. The second scalar field is neutral. The model has two conserved global charges with corresponding  
chemical potentials $\mu_1$ and $\mu_2$. We have introduced mass parameters $m_i>0$, self-couplings $\lambda_i>0$, and cross-species couplings $h$ and $G$. 
The dimensionful coupling constant $G$ gives rise to a derivative coupling, which models entrainment between the superfluid and the superconductor. In general, 
there are two independent entrainment couplings \cite{Haber:2015exa}, and in Eq.\ (\ref{Ux}) we have assumed them to be equal. This presents a simplification in particular 
for the temperature dependence which, in the given approximation, is absorbed in thermal masses $m_{i,T}$ and a thermal non-entrainment coupling $h_T$. In deriving
this temperature dependence, we have computed the excitation energies of the system. There are 6 modes in total: in the absence of condensation each complex field contributes 
two modes, plus two massless degrees of freedom from the gauge field; with condensation, these modes mix and, in the phase where neutral and charged condensates coexist, one massless Goldstone mode remains. For simplicity, we have applied a large-temperature approximation, although we shall employ the potential (\ref{Ux}) for all temperatures,
expecting at least a qualitatively correct result for the melting of the condensates.
In the derivation of the thermal potential we have also assumed that $G\mu_i^2,G m_i^2\ll 1$ 
(no assumption for the size of $GT^2$ was necessary). The expressions (\ref{mT}) are generalizations of well-known results for a single (charged) scalar field, 
see for instance Ref.\ \cite{Kapusta:1981aa}. They show for example that, even if 
one of the condensates vanishes, the other condensate knows about the presence of the second field through the thermal excitations. 

We are interested in the phase structure of the system for a homogeneous background magnetic field $\vec{H}$. To this end, we have to compare the Gibbs free energies of the various possible phases,
\be \label{FGibbs}
{\cal G}  = F - \frac{\vec{H}}{4\pi}\cdot\int d^3\vec{x} \, \vec{B}  \, .
\ee 
Taking into account the formation of magnetic flux tubes, we first need to compute the free energy of a single flux tube. 
This is done by solving the equations of motion for $\rho_1$, $\rho_2$, and $\vec{A}$ with the usual boundary conditions for a flux tube. This calculation has to be done numerically and is analogous to 
the calculation of Ref.\ \cite{Alford:2007np}, where a non-relativistic model has been employed at zero temperature. The free energy per unit length of a single flux tube with winding number $n$ turns out to be
\bea \label{Fn}
\frac{F_{\rm tube}}{L}&=&
\pi \rho_{01}^2 \int_0^\infty dr\,r\left\{\frac{n^2\kappa^2 a'^2}{r^2}+f_1'^2+f_1^2\frac{n^2(1-a)^2}{r^2}+\frac{(1-f_1^2)^2}{2}\right. \non[2ex]
&& \left. +x^2
\left[f_2'^2 + \frac{\lambda_2}{\lambda_1}x^2\frac{(1-f_2^2)^2}{2}\right]-\frac{h_T}{\lambda_1} x^2 (1-f_1^2)(1-f_2^2)-\Gamma x f_1f_2f_1'f_2'\right\} \, , 
\eea
where we have used cylindrical coordinates $\vec{x} = (R,\theta,z)$,  introduced the dimensionless radial coordinate $r=R/\xi$ with the coherence length of the charged condensate $\xi\equiv 1/(\sqrt{\lambda_1}\rho_{01})$, where $f_i(r)\equiv \rho_i(r)/\rho_{0i}$ with the homogeneous condensates in the COE phase $\rho_{0i}$, $a(r)\equiv rqA_\theta(r)/n$ with the gauge field $\vec{A}(r) = A_\theta(r)\vec{e}_\theta$. Moreover, $\kappa \equiv \sqrt{\lambda_1/(4\pi q^2)}$, 
$x\equiv \rho_{02}/\rho_{01}$, the dimensionless entrainment coupling is abbreviated by $\Gamma\equiv G\rho_{01}\rho_{02}$, and prime denotes derivative with respect to $r$.

We can then define the following three critical magnetic fields (abbreviation of the various phases as in Fig.\ \ref{fig:TcHc}).
\begin{itemize}
\item {\it Critical field $H_c$.} At $H_c$, the Gibbs free energies of the COE phase with all magnetic flux expelled ($\vec{B}=0$) and the SF phase (where $\vec{B}=\vec{H}$, assuming 
zero magnetization), are identical. Both phases are homogeneous, and thus we easily compute
\be \label{Hc}
H_c = \sqrt{2\pi \lambda_1}\,\rho_{01}^2\sqrt{1-\frac{h_T^2}{\lambda_1\lambda_2}} \, .
\ee  
In general, we also have to compute the critical magnetic field for the transition from 
the COE phase to the NOR phase, resulting in a slightly more complicated expression, which is appropriate for large temperatures, where  the NOR phase is preferred over the 
SF phase. In the numerical results discussed below, however, we shall compute the 
critical magnetic fields only at zero temperature, where Eq.\ (\ref{Hc}) is sufficient.

\item {\it Critical field $H_{c2}$.} Assuming a second-order phase transition between the COE phase with a flux tube lattice and the normal-conducting SF phase, 
$H_{c2}$ is the maximal magnetic field for which a nonzero charged condensate exists (becoming infinitesimally small just below $H_{c2}$). With the help of the linearized 
version of the equations of motion, this field can be computed \cite{tinkham,Sinha:2015bva}. Again, we find that the entrainment coupling only enters through the temperature effect, and we have a simple relation between $H_c$ and $H_{c2}$,
\be \label{Hc2}
H_{c2} = H_c \sqrt{2}\kappa\sqrt{1-\frac{h_T^2}{\lambda_1\lambda_2}} \, .
\ee
Consequently, $H_c$ and $H_{c2}$ are identical if $\lambda_1 -2\pi q^2 = h_T^2/\lambda_2$, and one can show that for magnetic fields just below $H_{c2}$ the Gibbs free energy of the superconducting phase is lower than that of the normal-conducting phase for $\lambda_1 -2\pi q^2 > h_T^2/\lambda_2$. So far, the results are a straightforward generalization of the textbook situation of an isolated 
superconductor, where $H_c$ and $H_{c2}$ intersect at $\kappa = 1/\sqrt{2}$. Only after computing $H_{c1}$ it becomes obvious that the generalization is less straightforward.
 
\item {\it Critical field $H_{c1}$.} At $H_{c1}$, it becomes favorable to place a single fluxtube (with winding number $n$) into the superconductor, resulting in a phase transition from the Meissner phase to the flux tube phase,
\be
H_{c1} = \frac{2q}{n}\frac{F_{\rm tube}}{L} \, ,
\ee
with the free energy for a single flux tube from Eq.\ (\ref{Fn}). This transition is continuous in the sense that the flux tube density is infinitesimally small just above $H_{c1}$. 
A discontinuous transition, where the flux tube density jumps from zero to a finite value, is possible if there is an attractive interaction at large distances between the flux tubes
(and short range repulsion) . 
Usually, in a single-component system, the change from repulsive to attractive interaction occurs exactly 
at $\kappa = 1/\sqrt{2}$, and a discontinuous onset of flux tubes never becomes relevant. This is different in the coupled system considered here, as we demonstrate below.  

\end{itemize}



In the current version, our model has 7 parameters, $m_1$, $m_2$, $\lambda_1$, $\lambda_2$, $h$, $G$, $q$,  
plus the 4 thermodynamic parameters $\mu_1$, $\mu_2$, $T$, and $H$. 
For simplicity, we set $G=0$ in the following phase diagrams. 
We set $q=2e\simeq 0.171$ (Gaussian units)
with the elementary charge $e$, having in mind a proton Cooper pair, and $m\equiv m_1=m_2$, having in mind neutrons and protons.   
Let us first discuss the $H=T=0$ case. If $h=0$, i.e., if the two fields are completely decoupled, it is obvious that coexistence of both condensates occurs if and only if $\mu_1>m$ and $\mu_2>m$. This coexistence region in parameter space is decreased for $h<0$ and increased for $h>0$
(boundedness of the potential requires $h<\sqrt{\lambda_1\lambda_2}$, allowing for arbitrary negative couplings, but imposing an upper limit for positive ones). 
In particular, for $h>0$ it is possible to achieve 
coexistence even if one (but not both) of the above conditions is violated, say $\mu_1<m$. And, for large negative values of the coupling, $|h|>\sqrt{\lambda_1\lambda_2}$, 
there is no coexistence possible for any $\mu_1$, $\mu_2$. 
Having this phase structure in mind, we choose $\mu_1,\mu_2>m$,  and discuss the phase structure for two values of the coupling, $\pm |h|$, with $|h|< \sqrt{\lambda_1\lambda_2}$.
For our purpose, it is most interesting to consider the plane spanned by the self-coupling of the fields $\lambda_1$, $\lambda_2$, because in this plane we expect a nontrivial structure regarding 
the change from type-I to type-II superconductivity, see above relation between $H_c$ and $H_{c2}$. 

\begin{figure} [t]
\begin{center}
\hbox{\includegraphics[width=0.5\textwidth]{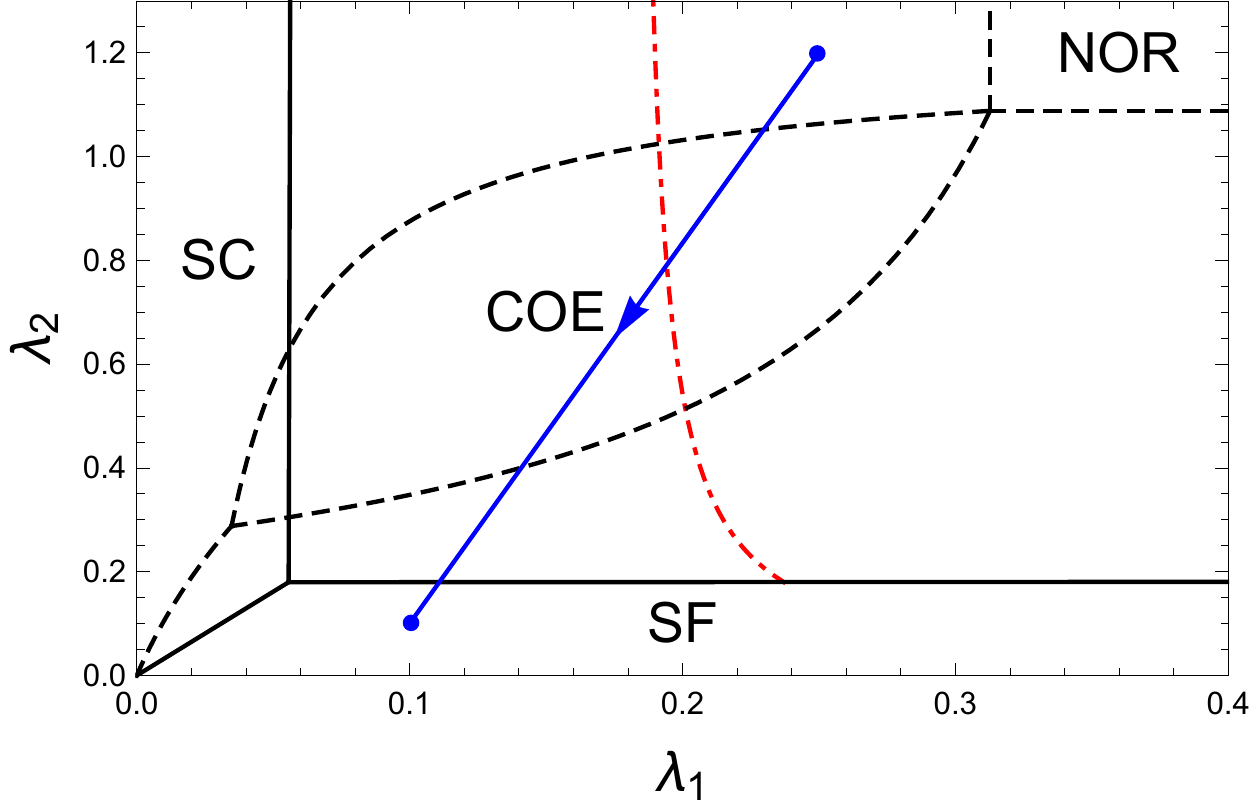}\includegraphics[width=0.5\textwidth]{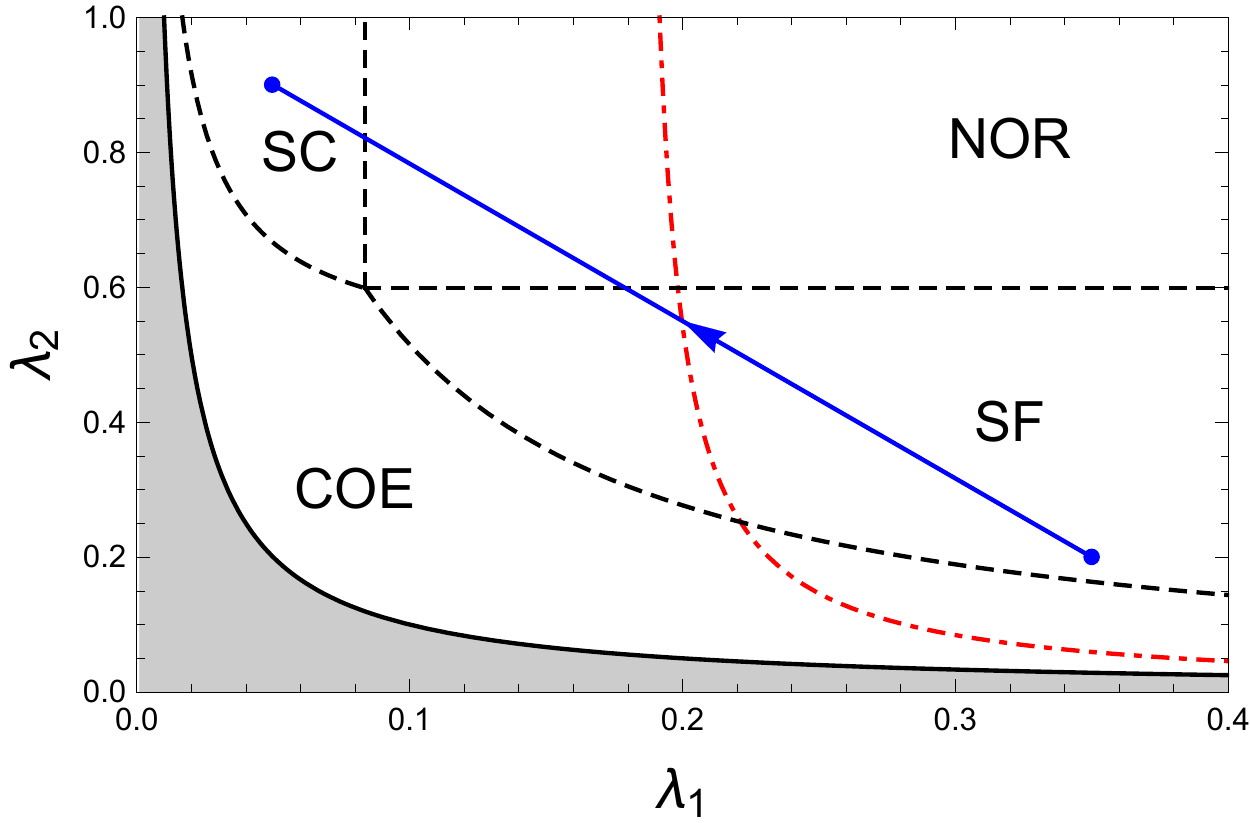}}
\caption{Phases in the $\lambda_1$-$\lambda_2$-plane for $h<0$ (left panel) and $h>0$ (right panel) at zero temperature [(black) solid curves] and nonzero temperature [(black) dashed curves], with the entrainment coupling set to zero, $G=0$. The shaded region in the right panel must be excluded because there the potential 
is unbounded. 
The (red) dashed-dotted curves -- only valid in the COE phase -- are given by $H_c = H_{c2}$ and do, in the absence of entrainment and in our approximation, not depend on $T$.  
The specific parameter sets are $q=0.17$, $\mu_1=1.5m$, $\mu_2=1.8m$ (both panels),
and $h=-0.1$, $T= 2.43m$ (left panel), $h=+0.1$, $T=3.5m$ (right panel). The (blue) lines ending in two dots are the paths taken for Fig. \ref{fig:TcHc}, parameterized by 
$\alpha\in[0,1]$ via $\lambda_i = \lambda_i^{\rm start}
+\alpha (\lambda_i^{\rm end} - \lambda_i^{\rm start})$ for $i=1,2$, and $(\lambda_1^{\rm start},\lambda_2^{\rm start})=(0.25,1.2)$ (left) and (0.35,0.2) (right), and
$(\lambda_1^{\rm end},\lambda_2^{\rm end})=(0.1,0.1)$ (left) and (0.05,0.9) (right).}
\label{fig:l1l2}
\end{center}
\end{figure}

We show the phase diagrams for positive and negative cross-coupling $h$ in Fig.\ \ref{fig:l1l2}, for zero temperature and one non-vanishing temperature, keeping $H=0$ for now.
We observe an 
asymmetric temperature effect concerning charged and neutral condensates:
the SC phase is disfavored more by temperature compared to the SF phase. This is a consequence of the increased thermal mass, 
see Eq.\ (\ref{mT}). The curve $H_c = H_{c2}$ divides the coexistence phase into two regions, expected to correspond to a superconductor of type-I (to the left of the curve) and of type-II (to the right of the curve). Before we switch on an external magnetic field, we recall the astrophysical context we are mainly interested in:
in the interior of a neutron star, there is only one parameter, the total baryon number density, which increases as we proceed into the core of the star, and all parameters in our model should eventually be functions 
of this density. Here we proceed simply by choosing a path parameterized by $\alpha\in [0,1]$  through the $\lambda_1$-$\lambda_2$ plane. There is of course some arbitrariness in choosing this path, but the purpose of the phase diagrams in Fig.\ \ref{fig:l1l2} was to show which scenarios are possible in general. We choose our path such that we 
cross from the type-II region for small $\alpha$ into the type-I region as $\alpha$
is increased, making $\alpha$ somewhat reminiscent of the baryon density. The results for the critical temperatures and zero-temperature critical magnetic fields  
are shown in Fig.\ \ref{fig:TcHc} and were already discussed in the introduction. Here we continue with a more detailed discussion of the type-I/type-II transition region, for which 
we have to compute the interaction between the flux tubes.

\section{Flux tube interaction and first-order flux tube onset}
\label{sec:inter}

We compute the flux tube interaction in the limit of a large flux tube separation, following Ref.\ \cite{Kramer:1971zza}, where the calculation has been done for a 
single-component superconductor, and Ref.\ \cite{Buckley:2003zf}, where a neutral component was added, but the interaction was only computed without entrainment
and approximated around the SU(2) symmetric case\ $\lambda_1=\lambda_2=h$. (As explained above, in our model this symmetric case is a singular point in parameter space 
because the tree-level potential becomes unbounded for all $h>\sqrt{\lambda_1\lambda_2}$).
The limit of large flux tube separations is particularly easy and can be treated semi-analytically 
because the lattice of flux tubes can be "patched" together from the single-flux tube solutions, which are cylindrically symmetric, plus small corrections. 

\begin{figure} [t]
\begin{center}
\hbox{\includegraphics[width=0.5\textwidth]{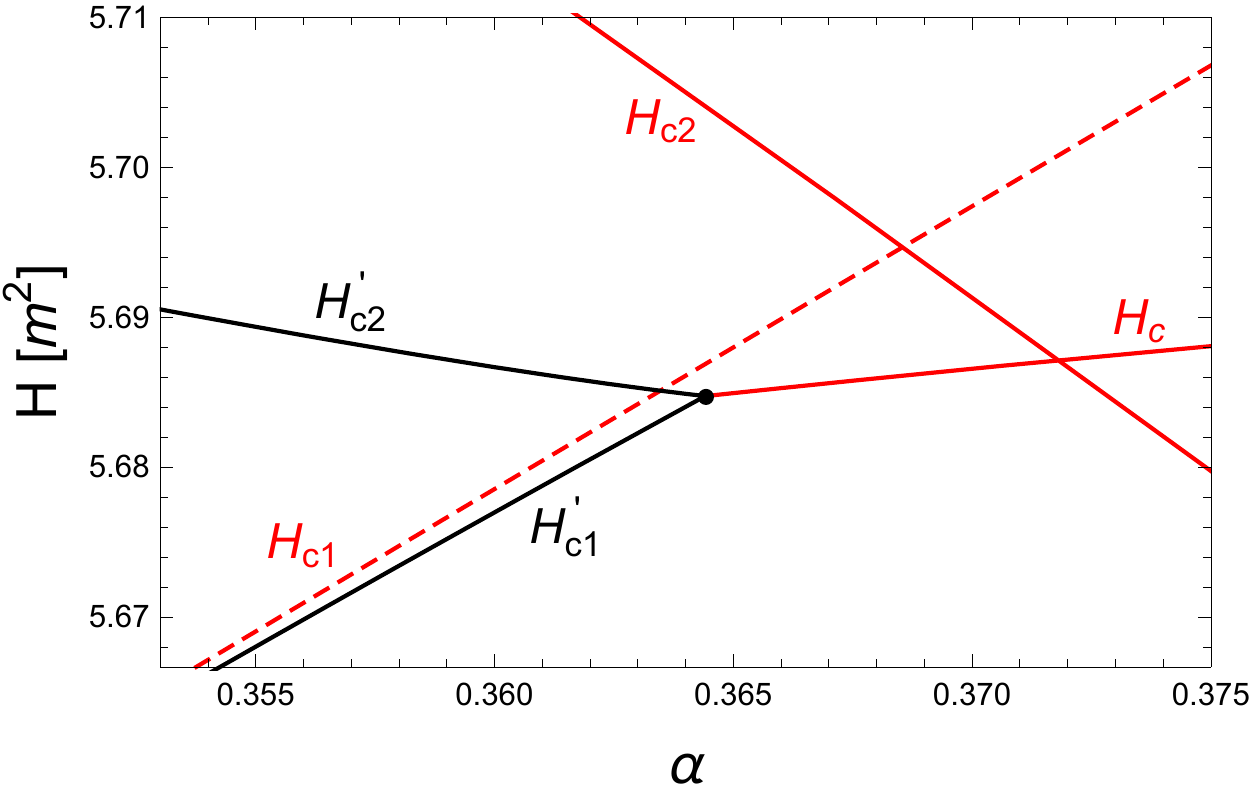}\includegraphics[width=0.5\textwidth]{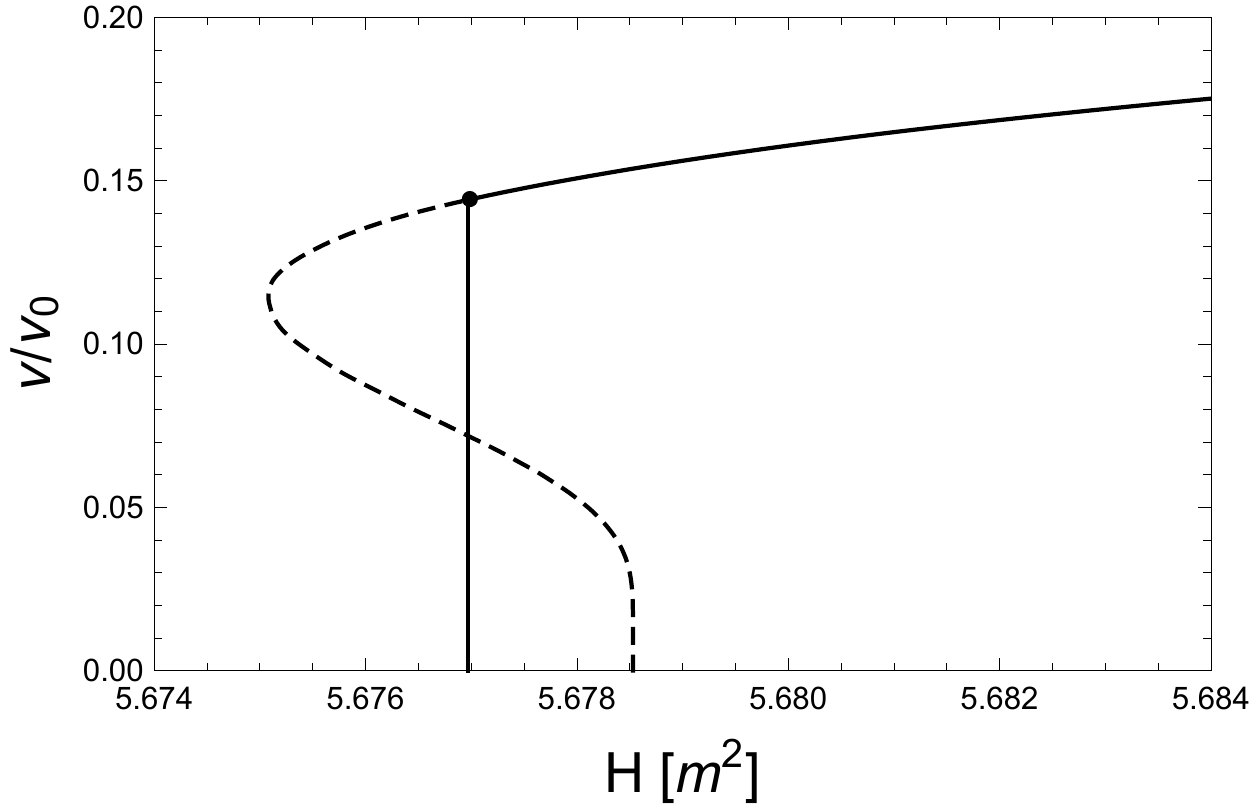}}
\caption{Left panel: zoom-in into the region where the three critical magnetic fields intersect, using the parameters from the left panels of Figs.\ \ref{fig:TcHc} and \ref{fig:l1l2}. 
$H_{c1}'$ and $H_{c2}'$ are first-order phase transitions, assuming a hexagonal flux tube lattice. The second-order onset with critical magnetic field $H_{c1}$ becomes 
discontinuous at $\alpha\simeq 0.292$ (beyond the scale of this plot, but shown in Fig.\ \ref{fig:TcHc}). Right panel: area density $\nu$ of the flux tubes as a function of 
$H$ for $\alpha = 0.360$ in units of $\nu_0 \equiv (\pi \xi^2)^{-1}$ (such that, roughly speaking, the flux tubes start to overlap 
 at $\nu/\nu_0 = 1$). The dashed segment of the curve is the unstable branch. }
\label{fig:intersect}
\end{center}
\end{figure}

We briefly sketch the idea of the approximation, the details of the calculation can be found in Ref.\ \cite{Haber:2017kth}. 
We consider two flux tubes in a distance $r_0$ and define the interaction energy by writing their total free energy as $F^{(1)}_{\rm tube}+ F^{(2)}_{\rm tube} +F_{\rm int}$, with 
$F^{(1)}_{\rm tube}$ and $F^{(2)}_{\rm tube}$ being the free energies of isolated flux tubes from Eq.\ (\ref{Fn}). We divide the total volume into two half-spaces separated by the plane perpendicular to and in the center of the line that connects the 
centers of the two flux tubes. One half-space is then the simplest version of a 
Wigner-Seitz cell. According to our assumption of a large separation $r_0$, the free energy in a certain half-space is given by the flux tube in that half-space plus a small correction from 
the other flux tube. This small correction can be formulated in terms of the solution for the linearized equations of motion, which is valid for the flux tube profile far away from its center. The correction to the free energy is a pure surface term if the first-order correction to the full equations of motion are kept to second order in the free energy. This reduces the problem to a surface integral over the plane separating the two Wigner-Seitz cells, and due to the simple geometry of the configuration we end up with 
a relatively compact result for the interaction free energy per unit length,
\bea \label{Fint}
\frac{F_{\rm int}}{L} &=& 2\rho_{01}^2 r_0 \int_{r_0/2}^\infty \frac{dr}{\sqrt{r^2-(r_0/2)^2}}\bigg\{\frac{\kappa^2 n^2 a'(1-a)}{r^2}-(1-f_1)f_1'-x^2(1-f_2)f_2'\non[2ex]
&&\hspace{4cm}+\frac{\Gamma x}{4}(f_1+f_2+f_1f_2-1)[(1-f_1)f_2'+(1-f_2)f_1']\bigg\} \, . 
\eea
We can find an analytical expression with the help of the asymptotic solutions of the equations of motion, 
\begin{subequations}
\bea
a(r) &\simeq & 1+ C rK_1(r/\kappa) \, , \label{Qasym} \allowdisplaybreaks \\[2ex]
f_1(r) &\simeq&  1+ D_+ \gamma_+K_0(\sqrt{\nu_+}r)+D_- \gamma_-K_0(\sqrt{\nu_-}r)  \, , \label{f1asym} \allowdisplaybreaks\\[2ex]
f_2(r) &\simeq& 1 + D_+ K_0(\sqrt{\nu_+}r)  +D_-K_0(\sqrt{\nu_-}r)\, , \label{f2asym}
\eea
\end{subequations}
where $K_p$ are the modified Bessel functions of second kind, where $\nu_\pm$ are the eigenvalues and $(\gamma_\pm,1)$ the eigenvectors of the matrix
\be
2\left(\begin{array}{cc} 1 & -\frac{\Gamma x}{2} \\[2ex] -\frac{\Gamma}{2x} & 1 \end{array}\right)^{-1} \left(\begin{array}{cc} 1 & -\frac{h_T}{\lambda_1}x^2  \\[2ex] -\frac{h_T}{\lambda_1} & \frac{\lambda_2}{\lambda_1}x^2 \end{array}\right) \, ,
\ee
which needs to be diagonalized to solve the linearized equations of motion. The coefficients $C$, $D_+$, $D_-$ have to be determined from the numerical solution. In the absence of entrainment, $\Gamma=0$, the interaction energy assumes the simple form 
\bea
\frac{F_{\rm int}}{2\rho_{01}^2\pi L} &=& \kappa^2 n^2 C^2 K_0(r_0/\kappa) - \left[D_+^2(\gamma_+^2+x^2)K_0(\sqrt{\nu_+}r_0)+D_-^2(\gamma_-^2+x^2)K_0(\sqrt{\nu_-}r_0)\right] \, .
\eea
From this expression we can determine the point at which the interaction at $r_0\to\infty$ changes from repulsive to attractive,
\bea\label{sign}
\frac{1}{\kappa^2} &=& 1+\frac{\lambda_2}{\lambda_1}x^2 -
\sqrt{\left(1-\frac{\lambda_2}{\lambda_1} x^2\right)^2 + \frac{4h^2 x^2}{\lambda_1^2}} = \frac{H_{c2}^2}{\kappa^2 H_c^2} \left[1-\frac{h^2}{\lambda_2^2 x^2}+{\cal O}\left(\frac{1}{x^4}\right)\right] \, .
\eea
Without coupling, $h=0$ (and $\lambda_2 x^2>\lambda_1$), the result becomes independent of $x=\rho_{02}/\rho_{01}$ and 
 we recover $\kappa^2=1/2$, i.e., 
in an isolated superconductor the flux tube interaction changes sign exactly where $H_c=H_{c1}=H_{c2}$. The coupling $h$ induces a correction to 
$\kappa^2=1/2$. As the expansion for large $x$ shows, for $x\to \infty$ this correction is the same as for the intersection point 
between $H_c$ and $H_{c2}$. 
This limit is interesting for neutron star applications because there the proton density is much smaller than the neutron density, which, translated into our formalism, 
means a large ratio of neutral to charged condensate, $x =\rho_{02}/\rho_{01} \gg 1$. In general, however, Eq.\ (\ref{sign}) shows that the flux tube interaction at large distances changes sign at a point different from $H_c=H_{c2}$. As a consequence, the phase transition to the flux tube phase can become discontinuous, which has been discussed in the literature in the context of multi-band superconductors \cite{PhysRevB.72.180502,PhysRevLett.105.067003,brandt2011attractive,2011PhRvB..84u4505L,2012arXiv1206.6786B,2014JPSJ...83c4701T}.
We denote the critical magnetic field for this discontinuous transition by $H_{c1}'$, and compute it as follows. 
For the flux tube lattice we employ the nearest neighbor approximation, which yields the Gibbs free energy per unit volume 
\be \label{FG1}
\frac{{\cal G}_{\rm lattice}}{V} \simeq U_{\rm COE} + \frac{n\nu}{2q}(H_{c1}-H) + \frac{t\nu}{2} \frac{F_{\rm int}}{L} \, , 
\ee
where $U_{\rm COE}$ is the free energy density of the homogeneous coexistence phase with all magnetic flux expelled, $t$ is the number of nearest neighbors,
and $\nu$ is the flux tube area density, related to $r_0$ by $\nu=s/r_0^2$, where $s$ depends on the lattice structure (for a hexagonal lattice, used in Fig.\ \ref{fig:intersect},
$t=6$ and $s=2/\sqrt{3}$). 
For the interaction term we use Eq.\ (\ref{Fint}) and assume the asymptotic values of the condensates to be identical to the homogeneous values in the Meissner phase, 
$\rho_{01}$ and $\rho_{02}$. Then, we minimize the Gibbs free energy with respect to $r_0$ (or, equivalently, with respect to $\nu$), insert the result back 
into  ${\cal G}_{\rm lattice}$ and determine the phase transition from the condition ${\cal G}_{\rm lattice}/V=U_{\rm COE}$.

After the flux tube lattice has been created at $H_{c1}'$, we may also ask at which critical magnetic field it is no longer favored over the normal-conducting phase. We denote this
critial magnetic field by $H_{c2}'$ and compute it from the condition ${\cal G}_{\rm lattice}/V=U_{\rm SF}-H^2/(8\pi)$. 
The results are shown in Fig.\ \ref{fig:intersect}. We can now identify the flux tube phase in the phase diagram, but have to keep in mind the range of validity of our approximation:
it is accurate only close to the point (\ref{sign}), where the second-order onset turns into a first-order onset. In the left panel of Fig.\ \ref{fig:intersect} our approximation becomes worse the 
further we follow the combined $H_{c1}'$, $H_{c2}'$ curve up right and then to the left along $H_{c2}'$. In
fact, in the left panel the incompleteness of our approach becomes manifest: $H_{c2}'$ suggests a first-order phase transition to the SF phase, but $H_{c2}$ is a lower 
boundary for that transition (excluding exotic scenarios where the flux tube phase disappears and then re-enters at a larger value of $H$). Hence, this 
phase diagram cannot be the final answer, not even qualitatively. A more complete, numerical calculation of the Gibbs free energy of the flux tube array is necessary. 
Such a calculation is beyond the scope of this work, but we show a conjectured phase structure in Fig.\ \ref{fig:HHH}, which has already been discussed in the introduction.

{\it Acknowledgments.} 
We thank Mark Alford, Nils Andersson, Carlos Lobo, and Andreas Windisch for helpful discussions and comments. We acknowledge support from the Austrian Science 
Fund (FWF) under project no.\ W1252 and from the {\mbox NewCompStar} network, COST Action MP1304. A.S.\ is supported by the Science \& Technology Facilities Council (STFC) in the form of an Ernest Rutherford Fellowship.

\bibliography{refs}

\end{document}